\newcommand{\Sup}{\uparrow}
\newcommand{\Sdn}{\downarrow}
\documentclass[twocolumn,showpacs,preprintnumbers,amsmath,amssymb]{revtex4}
\usepackage{graphicx}% Include figure files
\usepackage{dcolumn}% Align table columns on decimal point
\usepackage{bm}% bold math
\def\Dagcomp{^{\vphantom{\dagger}}}

\begin{document}

\title{Spin Hot Spots in vertically-coupled Few-electron Quantum Dots}
\author{Anjana Bagga, Pekka Pietil\"ainen$^{\dag}$
and Tapash Chakraborty}
\affiliation{Department of Physics and Astronomy,
University of Manitoba, Winnipeg, Canada R3T 2N2}
\affiliation{$^\dag$Department of Physical Sciences/Theoretical
Physics, P.O. Box 3000, FIN-90014 University of Oulu, Finland}
\date{\today}
\begin{abstract}
The effects of spin-orbit (SO) coupling arising from the confinement potential 
in single and two vertically-coupled quantum dots have been investigated. Our 
work indicates that a dot containing a single electron shows the lifting of the
degeneracy of dipole-allowed transitions at $B=0$ due to the SO coupling which
disappears for a dot containing two electrons. For coupled dots with one electron
in each dot, the optical spectra is not affected by the coupling and is the same
as the dot containing one electron. However, for the case of two coupled dots
where one partner dot has two interacting electrons while the other dot has only
one electron, a remarkable effect is observed where the oscillator strength of
two out of four dipole-allowed transition lines disappears as the distance between
the dots is decreased.
\end{abstract}
\maketitle

Interest in spin properties of semiconductor nanostructures has increased 
significantly in recent years due to the exciting possibility to manipulate 
it in solid state devices \cite{spintro,ohno}. The role of SO 
coupling in nanostructured systems is important in this context because this 
coupling would provide a means to influence the spin states via the orbital 
degrees of freedom \cite{nitta}. Quantum dots (QDs) are particularly promising 
systems for these studies as the electron spin states are very stable 
in these zero-dimensional systems and are measurable \cite{spin_measure}.

In quantum dots, spin hot spots are composed of two or more states that are 
degenerate in the absence of the SO coupling but the degeneracy is lifted due 
to the SO coupling \cite{spin_hot}. The importance of these hot spots lies 
in the fact that this lifting of degeneracy allows mixing of the spin-up and 
spin-down states and opens up the possibility for spin flip transitions in 
the presence of the SO coupling. Recent studies have indicated that for laterally 
coupled quantum dots, there is no contribution of the Bychkov-Rashba SO potential 
\cite{rashba} to the spin hot spots at zero magnetic field \cite{spin_hot}. 
We have explored the spin hot spots in vertically coupled quantum dots where 
they are readily identifiable and reflects interesting physical properties in 
the presence of SO coupling through the low-lying energy levels as well as in 
the dipole-allowed optical absorption spectra. To be specific, we have studied 
the effect of Bychkov-Rashba SO coupling in the optical transition spectra and 
the energy levels of two parabolic dots placed vertically and interacting only 
through the Coulomb interaction. Interestingly, the zero-field spin hot spots in our 
vertically-coupled QDs manifests in the dipole transition energies. For coupled
QDs with one dot containing two electrons while the other with a single electron,
the SO effect strongly depends on the interdot separation. It should be pointed 
out that a vast literature exists on the electronic properties of vertically-coupled 
quantum dots \cite{vertically}, but without any spin-orbit interaction included. 
The interest on the role of SO coupling in coupled quantum dots has reached its peak 
recently due to its importance in quantum information processing \cite{loss}.

We begin with the low-lying energy levels and the transition energies of a 
single electron in a vertically-coupled parabolic quantum dot \cite{comment,qdbook}
in the presence of a SO coupling. From the Dirac equation we know that 
whenever a spin-half particle with charge $q$ moves under the four-potential 
$(\vec A,\phi)$ the lowest-order relativistic correction leads to the SO
potential $V_{\mbox{\scriptsize SO}}$ of the form 
\begin{eqnarray*}
V_{\mbox{\scriptsize SO}}
&=&\frac{q\hbar}{4m^2c^2}\nabla\phi(\vec r)\cdot\vec\sigma\times
\left(\vec p-\frac qc\vec A(\vec r)\right) \\
&=&-\frac{q\hbar}{4m^2c^2}\vec E(\vec r)\cdot\vec\sigma\times
\left(\vec p-\frac qc\vec A\vec(r)\right). \\
\end{eqnarray*}
In quantum dots the electric field can arise, for example from the 
inversion asymmetry of the potential restricting the motion of the 
electrons (charge $e<0$ and effective mass $m^\ast$) to a two-dimensional 
plane \cite{chakra}. Then the electric field would be perpendicular 
to the plane of the motion. Furthermore, since we may well assume the field 
to be nearly homogenous within the range of the electron wave functions 
it can be replaced to a good approximation with its average value. It is 
also customary to collect all the parameters including the average of the 
electric field into a single coupling constant $\beta$ leading to the familiar
Bychkov-Rashba potential \cite{rashba}
$$ V_{\mbox{\scriptsize SOI}}=\beta\left[
\vec\sigma\times\left(\vec p-\frac ec\vec A(\vec r)\right)\right]_z, $$
for the SO coupling due to the inversion asymmetry.

Another source for the electric field in the quantum dot is of course
the potential $V_c$ that confines the electrons into the dot in the
two-dimensional plane. The field $\vec E$ now lies in the plane of motion and,
if the confinement is rotationally symmetric
it will be parallel to the radius vector $\vec r =(x,y)$. It is easy
to see that the SO coupling can now be written as \cite{voskoboynikov}
$$ V_{\mbox{\scriptsize SOC}}
=\sigma_z\frac{\alpha}{\hbar}\frac{dV_c(r)}{dr}
\left(p_\theta -\frac ec A_\theta(\vec r)\right), $$
where $p\Dagcomp_\theta$ and $A\Dagcomp_\theta$ stand for the angular 
components of the momentum and the vector potential, and where again we 
have combined most of the parameters to a constant $\alpha$. As compared 
to the $V_{\rm SOI}$ we see two essential differences in $V_{\rm SOC}$. 
Firstly, the coupling depends on the position, in particular in the case 
of the parabolic confinement $ V_c=\frac12 m^\ast\omega_0^2 r^2$
it will be proportional to $\omega_0^2r$. Secondly, it is diagonal
in spin space. From our point of view this latter property makes
it very attractive because it allows us to find analytic solutions
for the single-particle problem. It should be noted, however
that in order to see the effects arising from the confinement-%
induced SO coupling $V_{\mbox{\scriptsize SOC}}$, the confinement
itself must be rather large. Typically, $\hbar\omega_0$
must be of the order of 10--20\,meV \cite{voskoboynikov}. 

The Hamiltonian describing our coupled-dot system is given by
$${\cal H}=\sum_i{\cal H}_i^o + \sum_{i<j}
\frac{e^2}{\left[r_{ij}^2 + d^2 \right]^\frac12}$$	
where $d$ is interdot separation (in units of magnetic length $\ell_B$),
${\cal H}^o_i$ is the Hamiltonian governing a single electron 
confined in a parabolic quantum dot \cite{comment,qdbook} and is given by
\begin{equation}
\label{h1}
{\cal H}^o_i={\cal H}^o+{\cal H}^o_{\rm so}
\end{equation}
\begin{eqnarray*}
{\cal H}^o&=&\frac1{2m^*}\left(\vec{p}+\frac ec\vec{A}\right)^2+\frac12
\sigma_{z}\mu{B}gB + \frac12m^*\omega_o^2r^2 \\
{\cal H}^o_{so}&=&\sigma_z\alpha\dfrac{dV_c(r)}{dr} (k_\phi+
\dfrac e{2\hbar}Br) \\
  &=&\sigma_z\alpha(m\omega_o^2r)(k_\phi+\dfrac e{2\hbar}Br)
\end{eqnarray*}
where $\sigma_z$ is the Pauli $z$-matrix and $\alpha$ is the Bychkov-Rashba 
spin-orbit coupling parameter \cite{rashba}. It is to be noted that the first 
term of ${\cal H}^o_{so}$ is independent of the external magnetic field $B$. 
At $B=0$ it lifts the spin degeneracy because of the magnetic field orginating 
from the orbital motion of the electron in the presence of electric field 
coming from the confinement potential \cite{voskoboynikov}.

\begin{figure}[t]
\begin{center}
 \includegraphics[angle=0, width=.45\textwidth]{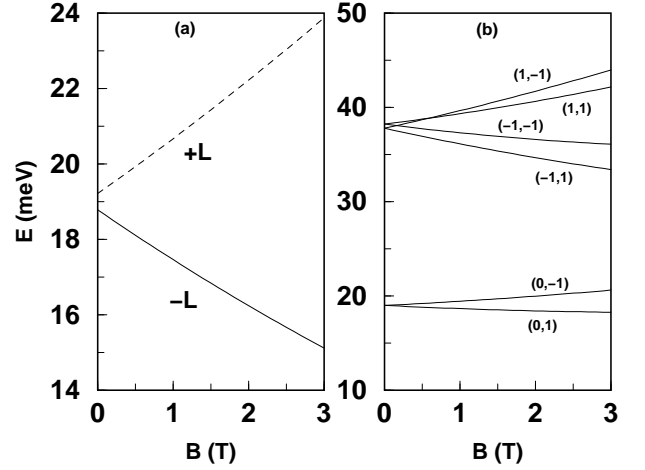}
\protect\caption{The (a) dipole transition energies and (b) a few low-lying 
energy levels of the single-electron energy spectrum of a single quantum dot 
in the presence of the SO coupling. The states are labelled as $(\ell,\sigma)$ 
where $\ell$ and $\sigma$ denote the orbital and spin quantum numbers respectively. 
The spin up and spin down projections are denoted by 1 and $-1$ respectively. 
At $B=0$, the degeneracy (without the SO coupling) is lifted. The states with 
antiparallel $\ell$ and spin are lower in energy.
}
\label{fig:one-electron}
\end{center}
\end{figure}

The energy eigenfunctions are then given by
$$ \psi_{n,\ell,\sigma}=\frac1{\sqrt2}\exp(i\ell\phi)R_{n,\ell,\sigma}(r) $$
where
\begin{eqnarray*}
R_{n,\ell,\sigma}(r)&=&\frac{\sqrt2}r\left[\frac{n!}{(n+\left|\ell\right|)!}
\right]^{\frac12} \exp\left(-\frac{r^2}{2r_{\sigma}^2}\right)\\
&\times&\left( \frac{r^2}{r_{\sigma}^2}\right)^\frac{\left|\ell\right|}2
L_n^{\left|\ell\right|} \left(\frac{r^2}{r_{\sigma}^2}\right)
\end{eqnarray*}
where $r_\sigma=(\hbar/m^*\Omega_\sigma)^{\frac12}$ and $L_{n}^{\left|
\ell\right|}$ is the generalized Laguerre polynomial.

\begin{figure}[t]
\begin{center}
 \includegraphics[angle=0, width=.42\textwidth]{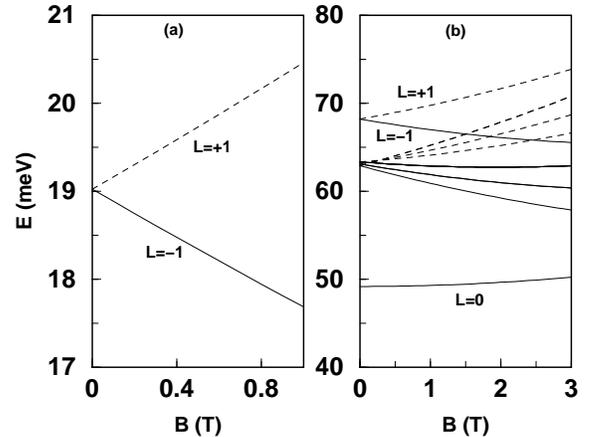}
%\vspace*{0.8cm}
\protect\caption{The (a) dipole-transition energy spectrum and (b) the few 
low-lying energy levels of a two-electron dot in the presence of the SO 
coupling. In (b) the lowest line represents the eigenvalues of the state 
$L=\ell_1+\ell_2=0, S=s_1+s_2=0$. The eigenvalues for the $L=1, S=1,0,-1$ 
states are drawn as dashed lines.
}
\label{fig:two-electrons}
\end{center}
\end{figure}

The corresponding electron energy levels are
\begin{equation}
\label{h3}
E_{n,\ell,\sigma}=\hbar\Omega_{\sigma}(2n+\left|\ell\right|+1)+\ell
\frac{\hbar\omega_c}2+\sigma\left[\frac{\mu_B}2gB+\ell\alpha m^*\omega_o^2\right]
\end{equation}
where
\begin{equation}
\label{h2}
\Omega_\sigma^2=\omega_o^2+\frac{\omega_c^2(B)}4+\sigma\alpha
\frac{m^*\omega_o^2}{\hbar}\omega_c,
\end{equation}
and $\omega_c$ is the cyclotron frequency. Clearly, the SO coupling influences 
$\Omega_{\sigma}$ [Eqs.~(\ref{h2}) and (\ref{h3})] which results in an 
increase of the energy of the up spin and an decrease in energy of the down 
spin. Another effect due to the SO coupling is from the last term of Eq.~(\ref{h3}) 
which is independent of an external magnetic field. As mentioned earler, at $B=0$ 
it lifts the spin degeneracy of the states with the same orbital momentum. The 
energy of the states with anti-parallel spin and $\ell$ is lowered while the 
states with the parallel $\ell$ and spin show an increase in energy at $B=0$.

\begin{figure}[t]
\begin{center}
 \includegraphics[angle=0, width=.42\textwidth]{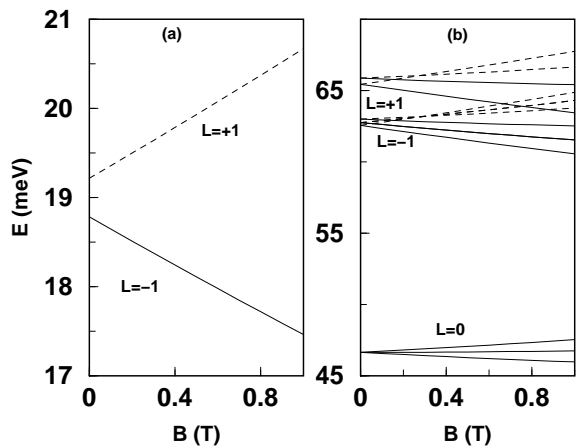}
\protect\caption{The (a) transition energy spectrum and (b) the few low-lying
energy levels of two vertically-coupled dots separated by $d=0.5\ell_B$
in the presence of the SO coupling. Each dot contains a single electron. 
In (b), the three lowest lines correspond to the eigenvalues for the states
$L=\ell_{D_1}+\ell_{D_2}=0, S=s_{D1}+s_{D2}=0,\pm1$. The eigenvalues for the 
$L=1$ states are drawn as dashed lines.
}
\label{fig:one_plus_one}
\end{center}
\end{figure}

The dipole-allowed transition energies of a single parabolic dot in 
the presence of the SO coupling are presented in Fig.~1(a)  while the 
energy eigenvalues are presented in Fig.~1(b). In all our calculations, 
the material parameters are for the InAs dots as listed in Ref.~\cite{voskoboynikov}. 
Transitions take place from the $\vert \ell=0,\Sup\rangle$ state to $\vert\ell=1,
\Sup\rangle$ and $\vert\ell=-1, \Sup\rangle$ states. These two transitions are not 
degenerate at $B=0$ because of the  SO coupling: $\vert-1\Sup\rangle$ has 
lower energy than that of $\vert 1\Sup\rangle$ [Fig.~1(b)]. It is interesting to 
note that this splitting between the $L=+1$ and $L=-1$ branches at $B=0$ disappears 
if the dot contains two electrons [Fig.~2(a)]. In the energy spectrum of a 
two-electron dot [Fig.~2(b)], the lowest line represents a two-electron state 
where the electrons occupy $\vert0\Sup\rangle$ and $\vert0\Sdn\rangle$ states. 
In order to understand the degeneracy of the $L=+1$ and $L=-1$ branches at $B=0$, 
let us look at the states involved in the transition to (say) the $L=+1$ branch. 
In this case, any of the two electrons occupying $\vert0\Sup\rangle$
and $\vert0\Sdn\rangle$ states can be excited: the electron is excited either 
from $\vert0\Sup\rangle$ to $\vert1\Sup\rangle$ or from $\vert0\Sdn\rangle$ to 
the $\vert1\Sdn\rangle$ state. Due to the SO interaction, the excitation energies 
are not degenerate at $B=0$. Since there are now two electrons, the Coulomb 
interaction between them mixes the $\vert 1\Sup\rangle$ and $\vert 1\Sdn\rangle$
states. The eigenstate is a combination of $\vert1\Sup\rangle$ and $\vert1\Sdn\rangle$ 
states due to the Coulomb interaction. Similarly, the $L=-1$ transition can occur 
in two ways: electron can either jump from $\vert 0\Sup\rangle$ to the 
$\vert-1\Sup\rangle$ state or from $\vert0\Sdn\rangle$ to the $\vert-1\Sdn\rangle$ 
state. Again, these two possibilities for $L=-1$ are not degenerate at $B=0$ but 
due to the Coulomb interaction the eigenstate is a combination of these two 
alternatives. However, the state $\vert-1\Sup\rangle$ is degenerate at $B=0$ 
with $\vert1\Sdn\rangle$ and so are the states $\vert-1\Sdn\rangle$
and $\vert1\Sup\rangle$ [Fig.~2(b)]. Therefore, we do not observe a splitting 
at $B=0$ between the $L=+1$ and $L=-1$ branches in the optical spectra of a dot 
containing two interacting electrons.

\begin{figure}[t]
\begin{center}
 \includegraphics[angle=0, width=.42\textwidth]{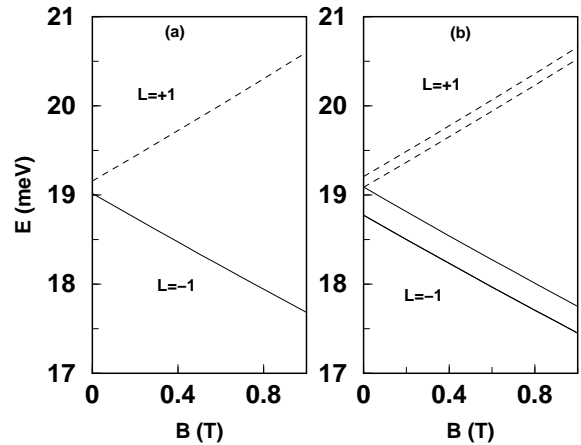}
\protect\caption{The transition energy spectrum of two vertically-coupled 
dots separated by (a) $d=0.5\ell_B$ and (b) $d=9.5\ell_B$, in the presence 
of the SO coupling. In this case, one dot confines a single electron while 
the other dot contains two interacting electrons.
}
\label{fig:two_plus_one}
\end{center}
\end{figure}

The results for two quantum dots each containing one electron and are 
vertically coupled (via only the Coulomb interaction) are displayed in Fig.~3. 
For a large separation of the dots (e.g., $d=9.5\ell_B$), the transition energies, 
quite expectedly, resemble those of a single dot with one electron [Fig.~1(a)]. 
As the dots are brought closer, the inter-dot interaction becomes stronger but 
that modifies only the energy spectrum, leaving the optical spectra unchanged 
[Fig. 3(a)].  Figure 3(b) shows the low-lying energy levels for two dots at a 
separation of $d=0.5 l_B$. An electron in the ground state $\left| 0 \uparrow 
\right\rangle$ in any of the two dots can make a transition to
$\left| 1 \uparrow \right\rangle$ state. There are now two states corresponding 
to $L=\ell_{D1}+\ell_{D2}=1$, $S=s_{D1}+s_{D2}=1$ where $\ell_{D1}$, $s_{D1}$ 
and $\ell_{D2}$, $s_{D2}$ represent the quantum numbers for the first and second 
dot respectively. When the inter-dot interaction is very weak, the two states 
($\ell_{D1}=1$, $\ell_{D2}=0$) and ($\ell_{D1}=0$, $\ell_{D2}=1$) are degenerate. 
However, as the separation between the dots is decreased, the Coulomb interaction 
lifts the degeneracy between the two eigenstates (which are linear combination of 
above two states). For $L=-1$, we similarly have two levels. However, the
oscillator strength of the lower state is nearly zero and the transition takes 
place only to the higher state. Therefore, there are only two lines (for $L=\pm 1$) 
in the optical spectra for all distances between the two dots [Fig. 3(a)].

The most interesting situation is found to occur when one dot has a single 
electron while the other dot contains two electrons. Figure 4 shows the 
transition energies of the two dots for two different values of the interdot 
separation: (a) $d=0.5 \ell_B$, and (b) $d=9.5\ell_B$. For a large separation 
of the dots $(d=9.5 \ell_B)$, there are four lines as a function of the 
magnetic field [Fig.~4(b)], whereas there are only two lines for a much 
smaller separation $(d=0.5\ell_B)$ [Fig.~4(a)]. This is because when the dots 
are far apart the uncoupled two-electron dot does not show the SO splitting 
between the $L=1$ and $L=-1$ branches at $B=0$ and as a result, there are 
two middle lines that are degenerate at $B=0$. The one-electron dot, on 
the other hand shows a splitting and results in the lowest line corresponding 
to the $L=-1$ branch while the uppermost line corresponds to the $L=1$ branch. 
Hence we have two $L=1$ branches and two $L=-1$ branches for the two dots. 
As the distance between the dots decreases, the excited $L=1$ state of the 
two dots are coupled by the Coulomb interaction. The transition probability 
for the lower state decreases while it increases for the higher state. At 
$d=0.5\ell_B$ the transition to the lower state is zero and hence we have 
only one line corresponding to the $L=1$ branch and similarly one line for 
the $L=-1$ branch [Fig.~4(a)]. The novel result we observe here is that, unlike 
in the case of two coupled dots with one electron in each dot, in the present 
system, for a large separation, the $L=1$ branch coming from the two-electron 
dot is {\it not} degenerate with the $L=1$ branch coming from the one-electron 
dot. Therefore, one could observe the disappearence of one of the two $L=1$ 
branches (same for the two $L=-1$ branches) as the distance between the two 
dots is decreased. It should be pointed out that the $L=1$ branch of the 
dot containing two electrons is not degenerate with the $L=1$ branch of the 
dot containing a single electron because the two-electron dot does not show 
the SO splitting between the $L=\pm1$ branches while the one-electron
dot does show the splitting.

In summary, we have investigated a single parabolic QD and two vertically 
coupled QDs, containing one or two (interacting) electrons, in the presence 
of the SO coupling. For single dots, the SO interaction that we considered 
to be arising from the confinement potential, lifts the degeneracy of the 
dipole-transition energies at $B = 0$ for a dot containing one electron.
The splitting disappears for a dot containing two electrons. The lifting 
of the degeneracy at $B=0$ is also observed for two coupled QDs, each 
containing a single electron. In case of two coupled dots where one partner 
dot has two interacting electrons while the other has only one
electron, the dipole transition energies show a remarkable dependence on the 
interdot separation. For a large separation, the spectra consist of four 
lines corresponding to a combination of one and two electron spectra. However, 
as the separation between the dots is decreased, the oscillator strength for 
the lower eigenstates of $L=+1$ and $L=-1$ decreases and
the optical spectra contains only two lines instead of four.

The work of T.C. has been supported by the Canada Research Chair
Program and the Canadian Foundation for Innovation (CFI) Grant.

\end{document}